\documentstyle[12pt]{article}
\begin{document}

\title{GRADIENT APPROACH TO
THE SPHALERON BARRIER }
\vspace{1.5truecm}
\author{
{\bf Guido Nolte}\\
Fachbereich Physik, Universit\"at Oldenburg, Postfach 2503\\
D-26111 Oldenburg, Germany
\and
{\bf Jutta Kunz}\\
Fachbereich Physik, Universit\"at Oldenburg, Postfach 2503\\
D-26111 Oldenburg, Germany\\
and\\
Instituut voor Theoretische Fysica, Rijksuniversiteit te Utrecht\\
NL-3508 TA Utrecht, The Netherlands}

\vspace{1.5truecm}


\maketitle
\vspace{1.0truecm}

\begin{abstract}
We apply the gradient approach to obtain
a path over the sphaleron barrier
and to demonstrate the fermionic level crossing phenomenon.
Neglecting the mixing angle dependence and
assuming that the fermions of a doublet are degenerate in mass
we employ spherically symmetric ans\"atze
for the fields.
The gradient path over the barrier is smooth,
even for large values of the Higgs boson mass
or of the fermion mass,
where the extremal energy path bifurcates.
\end{abstract}
\vfill
\noindent {Utrecht-Preprint THU-94/14} \hfill\break
\vfill\eject

\section{Introduction}

In 1976 't Hooft \cite{hooft} observed that
the standard model does not absolutely conserve
baryon and lepton number
due to the Adler-Bell-Jackiw anomaly.
The process 't Hooft considered was
spontaneous fermion number violation due to instanton
induced transitions.
Fermion number violating
tunneling transitions between
topologically distinct vacua might indeed
be observable at high energies at future accelerators
\cite{ring1,ssc}.

Manton considered
the possibility of fermion number
violation in the standard model
from another point of view \cite{man}.
Investigating the topological structure
of the configuration space of the Weinberg-Salam theory,
Manton showed that there are noncontractible
loops in configuration space, and
predicted the existence of a static, unstable solution
of the field equations, a sphaleron \cite{km},
representing the top of the energy barrier between
topologically distinct vacua.

At finite temperature this energy barrier between
topologically distinct vacua can be overcome
due to thermal fluctuations of the fields,
and fermion number violating
vacuum to vacuum transitions
involving changes of baryon and lepton number
can occur.
The rate for such baryon number violating processes
is largely determined by a Boltzmann factor,
containing the height of the barrier at a given
temperature and thus the energy of the sphaleron.
Baryon number violation in the standard model due to
such transitions over the barrier may be relevant
for the generation of the baryon asymmetry of the universe
\cite{krs,mcl1,mcl2,mcl3,kolb}.

While the barrier between topologically distinct vacua
is traversed,
the Chern-Simons number changes continuously
from $N_{\rm CS}=0$ in one vacuum sector
to $N_{\rm CS}\pm 1$ in the neighbouring vacuum sectors,
passing through the sphaleron at
$N_{CS}=\pm \frac{1}{2}$ \cite{akiba,bgkk}.
However, for large values of the Higgs boson mass,
energetically lower, asymmetric sphaleron solutions appear,
the bisphalerons \cite{kb1,yaffe}.
The minimum energy path over the barrier \cite{akiba}
then develops bifurcations \cite{bgkk},
indicating the need for another approach
to the sphaleron barrier,
which yields smooth paths.

As the barrier is traversed one occupied fermion level
crosses from the positive continuum
to the negative continuum or vice versa,
leading to the change in fermion number.
When considered in the background field approximation
this level crossing phenomenon
predicts the existence of a fermion zero mode
precisely at the top of the barrier, at the sphaleron
\cite{nohl,boguta,ring2}.
For massless fermions this zero mode
is known analytically \cite{nohl,boguta}.

Considering the minimum energy path over the barrier
\cite{akiba,bgkk}
the fermionic level crossing was demonstrated recently
in the background field approximation
under the assumption,
that the fermions of a doublet are degenerate in mass
\cite{kb2,dia,kb3}.
This assumption, violated in the standard model,
allows for spherically symmetric ans\"atze
for all of the fields, when the mixing angle dependence
is neglected (which is an excellent approximation
\cite{kkb1,kkb2}).
An analogous, but selfconsistent calculation \cite{nk1}
led to similar results.
However, for heavy fermions it led to
strongly deformed barriers, eventually giving rise
to bifurcations and to new sphalerons
\cite{nk1}.

Motivated by the catastrophes encountered
along the energy barrier
in the extremal energy path approach
for large Higgs boson or fermion masses
we here consider the gradient approach
to the sphaleron barrier.
In section 2 we briefly review
the Weinberg-Salam Lagrangian and the anomalous currents for vanishing
mixing angle and for degenerate fermion doublets.
In section 3 we discuss the sphaleron barriers.
We present the
radially symmetric ansatz for the boson fields
and obtain the energy functional.
We discuss the gradient approach,
putting special emphasis on the
question of the underlying metric.
We compare the barriers obtained with gradient approach
with those of the extremal energy path approach.
In section 4 we discuss the fermionic level crossing
along the gradient path over the barrier.
We present our conclusions in section 5.

\section{\bf Weinberg-Salam Lagrangian}

We consider the bosonic sector of the Weinberg-Salam theory
in the limit of vanishing mixing angle.
In this limit the U(1) field
decouples and can consistently be set to zero
\begin{equation}
{\cal L}_{\rm b} = -\frac{1}{4} F_{\mu\nu}^a F^{\mu\nu,a}
+ (D_\mu \Phi)^{\dagger} (D^\mu \Phi)
- \lambda (\Phi^{\dagger} \Phi - \frac{1}{2}v^2 )^2
\   \end{equation}
with the SU(2)$_L$ field strength tensor
\begin{equation}
F_{\mu\nu}^a=\partial_\mu V_\nu^a-\partial_\nu V_\mu^a
            + g \epsilon^{abc} V_\mu^b V_\nu^c
\ , \end{equation}
and the covariant derivative for the Higgs field
\begin{equation}
D_{\mu} \Phi = \Bigl(\partial_{\mu}
             -\frac{1}{2}ig \tau^a V_{\mu}^a \Bigr)\Phi
\ . \end{equation}
The ${\rm SU(2)_L}$
gauge symmetry is spontaneously broken
due to the non-vanishing vacuum expectation
value $v$ of the Higgs field
\begin{equation}
    \langle \Phi \rangle = \frac{v}{\sqrt2}
    \left( \begin{array}{c} 0\\1  \end{array} \right)
\ , \end{equation}
leading to the boson masses
\begin{equation}
    M_W = M_Z =\frac{1}{2} g v \ , \ \ \ \ \ \
    M_H = v \sqrt{2 \lambda}
\ . \end{equation}
We employ the values $M_W=80 \ {\rm GeV}$, $g=0.65$.

For vanishing mixing angle,
considering only fermion doublets degenerate in mass,
the fermion Lagrangian reads
\begin{eqnarray}
{\cal L}_{\rm f} & = &
   \bar q_L i \gamma^\mu D_\mu q_L
 + \bar q_R i \gamma^\mu \partial_\mu q_R
   \nonumber \\
           & - & f^{(q)} \bar q_L
           (\tilde \Phi u_R + \Phi d_R)
               - f^{(q)} (\bar d_R \Phi^\dagger
                     +\bar u_R \tilde \Phi^\dagger)
           q_L
\ , \end{eqnarray}
where $q_L$ denotes the lefthanded doublet
$(u_L,d_L)$,
while $q_R$ abbreviates the righthanded singlets
$(u_R,d_R)$,
with covariant derivative
\begin{equation}
D_\mu q_L = \Bigl(\partial_{\mu}
             -\frac{1}{2}ig \tau^a V_{\mu}^a \Bigr) q_L
\ , \end{equation}
and with $\tilde \Phi = i \tau_2 \Phi^*$.
The fermion mass is given by
\begin{equation}
M_F=\frac{1}{\sqrt{2}}f^{(q)} v
\ . \end{equation}

Due to the U(1) anomaly
baryon number and lepton number are not conserved
\begin{equation}
\partial^\mu j_\mu^{\rm B,L}=-f_{\rm g} \partial^\mu K_\mu
\ ,  \end{equation}
where
\begin{equation}
 K_\mu=\frac{g^2}{16\pi^2}\varepsilon_{\mu\nu\rho\sigma} {\rm Tr}(
 {\cal F}^{\nu\rho}
 {\cal V}^\sigma
 + \frac{2}{3} i g {\cal V}^\nu {\cal V}^\rho {\cal V}^\sigma )
\   \end{equation}
(${\cal F}_{\nu\rho} = 1/2 \tau^i F^i_{\nu\rho}$,
${\cal V}_\sigma = 1/2 \tau^i V^i_\sigma$)
is the Chern-Simons current and $f_{\rm g}$ is the number of
generations.
In the unitary gauge the topological baryon number $Q_{\rm B}$,
carried by a configuration, is determined
by its Chern-Simons number $N_{\rm CS}$,
\begin{equation}
N_{\rm CS} = \int d^3r K^0
\ . \end{equation}
For the vacua the Chern-Simons number is identical to the
integer winding number,
while the sphaleron at the top of the barrier carries
half integer Chern-Simons number \cite{km}.

\section{Barriers}

Approximations to the sphaleron barrier can be obtained by
constructing families of field configurations
for the  gauge and Higgs boson fields,
which interpolate smoothly from one vacuum sector to a neighbouring
one.
The minimum of all maximum energy configurations
encountered along such vacuum to vacuum paths
represents the sphaleron \cite{km} or,
for large Higgs boson masses, the bisphaleron \cite{kb1,yaffe}.
Applying the gradient approach, we construct the sphaleron
barrier for various Higgs boson masses.
We compare these barriers to the ones obtained
by constructing the extremal energy path \cite{akiba,bgkk}.

\subsection{\bf Energy Functional}

In the limit of vanishing mixing angle
the gauge and Higgs boson fields can be parametrized by a
spherically symmetric ansatz
given by \cite{dhn}
\begin{eqnarray}
  V_i^a & = & \frac{1-f_A(r)}{gr} \varepsilon_{aij}\hat r_j
  + \frac{f_B(r)}{gr} (\delta_{ia}-\hat r_i \hat r_a)
  + \frac{f_C(r)}{gr} \hat r_i \hat r_a  \ ,
 \label{ans1}\\
  V_0^a & = & 0\ , \\
    \Phi & = & \frac{v}{\sqrt {2}}
  \Bigl(H(r) + i \vec \tau \cdot \hat r K(r)\Bigr)
    \left( \begin{array}{c} 0\\1  \end{array} \right)
\label{ans2}
\ , \end{eqnarray}
which involves the five radial functions
$f_A(r)$, $f_B(r)$, $f_C(r)$,
$H(r)$ and $K(r)$.

This ansatz is form-invariant under
spherically symmetric gauge transformations with
the SU(2) matrix
\begin{equation}
U(\vec r\,)=\exp(i\frac{\Theta(r)}{2} \vec\tau \cdot \hat r)
\ . \label{gauge2} \end{equation}
The functions then transform as
\begin{eqnarray}
f_A+if_B &\longrightarrow& \exp(i\Theta)(f_A+if_B) \ , \nonumber\\
f_C &\longrightarrow& f_C+r\Theta'  \ , \nonumber\\
H+iK &\longrightarrow&  \exp(i\frac{\Theta}{2})(H+iK) \ .
\label{gauge}
\end{eqnarray}

The ansatz, Eqs.~(\ref{ans1})-(\ref{ans2}), leads to the
bosonic energy functional
\begin{eqnarray}
 E_{\rm b} & = & \frac{4\pi M_W}{g^2} \int^{\infty}_0 dx
         \Bigl[  \frac{1}{2x^2} (f^2_A + f^2_B  -1)^2
         + (f'_A + \frac{f_Bf_C}{x})^2
         + (f'_B - \frac{f_Af_C}{x})^2
\nonumber \\
   & + & (K^2+H^2) (1+f_A^2+f^2_B +
         \frac{f_C^2}{2})+2f_A (K^2-H^2) - 4f_B H K
\nonumber \\
   & - & 2xf_C (K'H - KH') + 2x^2(H'^2+K'^2)
      + \epsilon x^2 (H^2+K^2 -1)^2
    \Bigr]
\ , \label{engy} \end{eqnarray}
where
\begin{equation}
x=M_Wr \nonumber
\   \end{equation}
is a dimensionless coordinate,
the prime denotes differentiation with respect to $x$, and
\begin{equation}
\epsilon = \frac{4\lambda}{g^2}
         = \frac{1}{2}\Bigl( \frac{M_H}{M_W} \Bigr)^2 \nonumber
\ . \end{equation}
This energy functional has a residual U(1) gauge invariance
with respect to the gauge transformation, Eq.~(\ref{gauge}).

The Chern-Simons number of a given configuration is
\begin{equation}
N_{\rm CS} = \frac{1}{2\pi} \int^{\infty}_0 \ dx \Bigl[
(f_A^2+f^2_B) (\frac{f_C}{x} - \theta') -
(\frac{f_C}{x} - \Theta') -
\Bigl(\sqrt {(f_A^2+f^2_B)}\
\sin (\theta - \Theta)\Bigr)'\ \Bigr]
\ , \label{ncs} \end{equation}
where
\begin{equation}
\theta(x) = {\rm arctan}({f_B(x)}/{f_A(x)})
\ . \end{equation}
The function $\Theta(x)$ is an arbitrary radial function,
associated with the U(1) gauge transformation, Eq.~(\ref{gauge}).
 From the expression (\ref{ncs})
the Chern-Simons number is readily obtained in an arbitrary gauge.

In the radial gauge, where $f_C=0$,
the spatial part of the Chern-Simons current
contributes to the topological baryon number.
One therefore has to rotate to the unitary gauge,
where only the Chern-Simons number
determines the topological baryon number.
The corresponding gauge transformation
involves the function $\Theta(x)$,
which satisfies $\Theta(0)=0$ and
$\Theta(\infty)=\theta(\infty)$.
This leads to the Chern-Simons number
\begin{equation}
N_{\rm CS}=
\frac{1}{2\pi}\int_0^{\infty}dx
(f_B f_A'-f_A f_B') +\frac{\theta(\infty)}{2\pi}
\ . \end{equation}

The energy functional, Eq.~(\ref{engy}), possesses non-trivial
extrema.
The sphaleron \cite{km} with $N_{\rm CS}=1/2$
exists for all Higgs boson masses.
For large Higgs boson masses the
energetically lower, asymmetric bisphalerons
bifurcate from the sphaleron \cite{kb1,yaffe}.

\subsection{Gradient Approach}

Let us now consider the energy barriers,
associated with the sphaleron and,
for large Higgs boson masses, the bisphalerons.
For instance, the minimum energy path
over the sphaleron barrier is obtained from the functional \cite{akiba}
\begin{equation}
W=E_{\rm b}  + \frac{8 \pi^2 M_W} {g^2} \xi N_{\rm CS}
\ , \label{mep} \end{equation}
where $\xi$ is a dimensionless Lagrange multiplier \cite{akiba}.
This path is satisfactory for small Higgs boson masses \cite{akiba}.
For large Higgs boson masses, however, variation of the
functional, Eq.~(\ref{mep}), leads to an extremal energy path,
which does not culminate at the bi\-sphaleron,
but has a spike in the vicinity of the bisphaleron
and culminates at the sphaleron \cite{bgkk}.
This is in clear contrast to expectation and
to other paths constructed in more or less ad hoc fashion
\cite{klink,kb3}.
Similarly, when the path is constructed in the presence of
fermions with large masses, bifurcations arise along the path
in the vicinity of the sphaleron \cite{nk1}.

These catastrophic features of the extremal energy path
are artifacts of this approach,
indicating the need for another systematic approach
to obtain the sphaleron barrier.

\subsubsection{Explicit gradient formalism}

Let us therefore consider the gradient approach
as an alternative approach to the spha\-le\-ron barrier.
Starting at the sphaleron or bisphaleron
we are looking for the steepest path,
connecting the top of the barrier with the vacua on both sides.

The direction of steepest descent
at the top of the barrier is determined
by the negative mode of the sphaleron or bisphaleron
\cite{yaffe,bk1}.
Away from the top of the barrier,
the gradient of the energy functional determines
the direction of steepest descent.
Thus for a given configuration $\tilde f$
the neighbouring configuration $f=\tilde f+\delta f$
along the path of steepest descent
is obtained by choosing $\delta f$
proportional to the gradient of the energy functional.
Explicitly, when $f$ denotes a set of functions $f_i$,
we obtain $\delta f$ from the
functional derivative of the bosonic energy functional $E_{\rm b}$
according to
\begin{equation}
\delta f_i=\alpha \Biggl.\left[
\frac{\partial {\cal E}(f,f')}{\partial f_i(x)}-
\left(\frac{\partial {\cal E}(f,f')}{\partial f_i'(x)}\right)' \
\right]\Biggr|_{
f=\tilde{f},f'=\tilde{f}'}
\ , \label{df} \end{equation}
with
\begin{equation}
E_{\rm b} = \int dx {\cal E}(f_i, f_i')
\ , \end{equation}
and $\alpha$ is a small negative number.

{\it Metric}

The notion ``steep'' always refers to a metric on the configuration
space.
Therefore, to unambiguously define the gradient approach,
we need to specify a metric on the configuration space.
A natural choice for this metric is provided by
the kinetic energy term of the Lagrangian \cite{met}.
For the spherically symmetric ansatz, Eqs.~(\ref{ans1})-(\ref{ans2}),
the effective mass
\cite{banks1,banks2,bitar1,bitar2,noble,hsu}
of the gauge-Higgs-system reads
\begin{equation}
m(\lambda)=\frac{8\pi}{g^2M_W}
 \int dx \left[\left(\frac{df_A}{d\lambda}\right)^2+
 \left(\frac{df_B}{d\lambda}\right)^2
+\frac{1}{2}\left(\frac{df_C}{d\lambda}\right)^2
+2x^2\left(\frac{dH}{d\lambda}\right)^2
+2x^2\left(\frac{dK}{d\lambda}\right)^2\right]
\ , \end{equation}
where $\lambda$ is an arbitrary path parameter.

Accordingly we define a distance $\hat{d}$ of two
configurations
$f=(f_A,f_B,f_C,H,K)$ and
$\tilde f = (\tilde{f}_A,\tilde{f}_B,
\tilde{f}_C,\tilde{H},\tilde{K})$,
taken at the ``times'' $\lambda$ and $\tilde \lambda$, as
\begin{eqnarray}
 \hat{d}^2(f,\tilde{f})&=&
\frac{16\pi}{g^2}\int dx \Bigg[
(f_A-\tilde{f_A})^2
+(f_B-\tilde{f_B})^2
+\frac{1}{2}(f_C-\tilde{f_C})^2
\nonumber \\
&&
\phantom{\frac{16\pi}{g^2}\int dx \Bigg[}
+2x^2(H-\tilde{H})^2
+2x^2(K-\tilde{K})^2\Bigg]
\ . \label{dis2} \end{eqnarray}
Since the gradient formalism
assumes a Euclidian metric with equal weight
for all indices
(i.~e.~space points and function indices),
corresponding to a distance $\hat{d}$
\begin{equation}
\hat{d}^2(f,\tilde{f})=\frac{16\pi}{g^2}
\sum_{i}\int_0^{\infty}dx \left(
f_i-\tilde{f}_i\right)^2
\ , \label{dis1} \end{equation}
we have the relations
$f_1(x)=f_A(x)$, $f_2(x)=f_B(x)$, $f_3(x)=\frac{1}{\sqrt{2}}f_C(x)$,
$f_4(x)=\sqrt{2}xH(x)$ and  $f_5(x)=\sqrt{2}xK(x)$.

{\it Equations}

Assuming the old configuration
$\tilde f = (\tilde{f}_A,\tilde{f}_B,
\tilde{f}_C,\tilde{H},\tilde{K})$
is in the radial gauge, $\tilde{f}_C=0$,
the new configuration $f=(f_A,f_B,f_C,H,K)$
will in general not be in the radial gauge, $f_C \neq 0$.
Denoting
$\delta f=f-\tilde{f}$,
we find the set of equations
\begin{equation}
\delta f_A=2\alpha \Bigl[ \frac{\tilde{f}_A}{x^2}(\tilde{f}_A^2
+\tilde{f}_B^2-1)-\tilde{f}_A''
+\tilde{f}_A(\tilde{K}^2+\tilde{H}^2)+\tilde{K}^2-\tilde{H}^2
\Bigr]
\ , \end{equation}
\begin{equation}
\delta f_B=2\alpha \Bigl[ \frac{\tilde{f}_B}{x^2}(\tilde{f}_A^2
+\tilde{f}_B^2-1)-\tilde{f}_B''
+\tilde{f}_B(\tilde{K}^2+\tilde{H}^2)-2\tilde{H} \tilde{K}
\Bigr]\
\ , \end{equation}
\begin{equation}
\delta f_C=\frac{4\alpha}{x} \Bigl[ \tilde{f}_A'\tilde{f}_B
-\tilde{f}_B'\tilde{f}_A
-x^2(\tilde{K}'\tilde{H}-\tilde{H}'\tilde{K})
\Bigr]
\ , \end{equation}
\begin{equation}
\delta H=\frac{\alpha}{x^2} \Bigl[
\tilde{H}(1+\tilde{f}_A^2+\tilde{f}_B^2)-2\tilde{f}_A \tilde{H}
 -2 \tilde{f}_B \tilde{K}+2\tilde{H} \epsilon x^2(\tilde{H}^2+\tilde{K}^2-1)
-(2x^2\tilde{H}')'
\Bigr]
\ , \end{equation}
\begin{equation}
\delta K=\frac{\alpha}{x^2} \Bigl[
\tilde{K}(1+\tilde{f}_A^2+\tilde{f}_B^2)+2\tilde{f}_A \tilde{K}
 -2 \tilde{f}_B \tilde{H}+2\tilde{K} \epsilon x^2(\tilde{H}^2+\tilde{K}^2-1)
-(2x^2\tilde{K}')'
\Bigr]
\ . \end{equation}

To obtain the new configuration
in the radial gauge,
$\hat f=(\hat f_A,\hat f_B,\hat f_C,\hat H,\hat K)$,
we perform a gauge transformation, Eq.~(\ref{gauge}),
of $f=(f_A,f_B,f_C,H,K)$
with the gauge function $\Phi$
determined by $\hat f_C = 0 = f_C+x \Phi'$.

\subsubsection{Gradient formalism with constraint}

Let us now consider a modified gradient formalism,
which is equivalent to the explicit gradient approach
in the limit $\hat d(f,\tilde f) \rightarrow 0$.
In the explicit gradient formalism
the gradient is taken at the `old' configuration $\tilde{f}$.
In the modified gradient formalism
we take the gradient at the `new' configuration
$f=\tilde{f}+\delta f$.
Then Eq.~(\ref{df}) is a set of differential equations for
the functions $f_i$,
which can also be obtained by variation of the functional $W$
\begin{equation}
W(f)=E_{\rm b}(f)+\frac{1}{4}\xi M_W \hat d^2(f,\tilde{f})
\ , \label{w} \end{equation}
where a constraint is imposed on the distance $\hat{d}(f,\tilde{f})$,
and $\xi$ is a Lagrange multiplier
inverse proportional to $\alpha$.

{\it Choice of gauge}

As before, starting from a configuration in the radial gauge,
$\tilde f_C=0$, the neighbouring configuration
will in general not be in the radial gauge, $f_C\neq 0$.
To be able to keep the radial gauge throughout,
we define a new distance $d(f,\tilde f)$
\begin{eqnarray}
&&d (f,\tilde{f})=\min_{\Phi}\hat{d}(f,\tilde f_{-\Phi})=\nonumber \\
&&\Bigg\{\frac{16\pi}{g^2}\min_{\Phi}\int dx \left[
\left(f_A-(\tilde{f_A}\cos(\Phi)
+\tilde{f_B}\sin(\Phi))\right)^2
+\left(f_B-(\tilde{f_B}\cos(\Phi)
-\tilde{f_A}\sin(\Phi))\right)^2 \right.\nonumber\\
&& +2x^2\left( H-(\tilde{H}\cos(\Phi/2)
+\tilde{K}\sin(\Phi/2))\right)^2
+2x^2\left(K-(\tilde{K}\cos(\Phi/2)
-\tilde{H}\sin(\Phi/2))\right)^2\nonumber\\
&&\left.
+\frac{1}{2}\left(f_C-\tilde{f_C}+x\Phi'\right)^2\right]\Bigg\}^{1/2}
\ , \label{dis3} \end{eqnarray}
where $\tilde f_{\Phi}$ is obtained by
gauge transforming $\tilde f$ with the
gauge function $\Phi$, which minimizes the distance (\ref{dis3}).
In contrast with the distance, Eq.~(\ref{dis2}),
the new distance is gauge invariant
under independent gauge transformations of $f$ and $\tilde f$.
(Denoting the configuration space by $C$,
this is a proper metric on the projective
space of gauge orbits $C/\sim$,
where the equivalence relation $\sim$ identifies
configurations which are connected by radially symmetric
gauge transformations, Eq.~(\ref{gauge}) \cite{fub}.)

{\it Equations}

Variation of the functional $W$, Eq.~(\ref{w}),
with distance $\hat d(f,\tilde f)$ replaced by
$d(f,\tilde f)$
leads in the gauge $f_C=\tilde f_C=0$
for the gauge and Higgs field functions to the set of equations
\begin{equation}
f_A''=\frac{f_A}{x^2}(f_A^2+f_B^2-1)+f_A(K^2+H^2)
+K^2-H^2
+\xi \left(f_A-(\tilde{f}_A\cos(\Phi)+\tilde{f}_B\sin(\Phi))\right)
\ , \end{equation}
\begin{equation}
f_B''=\frac{f_B}{x^2}(f_A^2+f_B^2-1)+f_B(K^2+H^2)
-2HK
+\xi \left(f_B-(\tilde{f}_B\cos(\Phi)-\tilde{f}_A\sin(\Phi))\right)
\ , \end{equation}
\begin{eqnarray}
H''&=& -\frac{2}{x}H'
+\frac{H}{2x^2}\Bigl((1-f_A)^2+f_B^2\Bigr) -\frac{K}{x^2}f_B
+(H^2+K^2-1)H
\nonumber \\
&+&\xi \left(H-(\tilde{H}\cos(\Phi/2)+\tilde{K}\sin(\Phi/2))\right)
\ , \end{eqnarray}
\begin{eqnarray}
K''&=& -\frac{2}{x}K'
+\frac{K}{2x^2}\Bigl((1+f_A)^2+f_B^2\Bigr) -\frac{H}{x^2}f_B
+\epsilon (H^2+K^2-1)K
\nonumber \\
&+&\xi \left(K-(\tilde{K}\cos(\Phi/2)-\tilde{H}\sin(\Phi/2))\right)
\ , \end{eqnarray}
and to an additional equation for $\Phi$
\begin{eqnarray}
\Phi''&=&-\frac{2}{x}\Phi'+\frac{2}{x^2}
\sin(\Phi)(f_A\tilde{f}_A+f_B\tilde{f}_B)
+\frac{2}{x^2}\cos(\Phi)(f_B\tilde{f}_A-f_A\tilde{f}_B) \nonumber \\
 &  &+2\sin(\Phi/2)(H\tilde{H}+K\tilde{K})
+2\cos(\Phi/2)(K\tilde{H}-H\tilde{K})
\ . \label{phi} \end{eqnarray}

{\it Boundary conditions}

The boundary conditions are chosen such that both the energy density
and the energy are finite.
At the origin the gauge and Higgs field functions satisfy
the boundary conditions
\begin{equation}
f_A(0)-1=f_B(0)=H'(0)=K(0)=0
\ . \end{equation}
At infinity the gauge and Higgs field functions lie on a circle
\begin{equation}
f_A(\infty)+if_B(\infty)=\exp(i\theta(\infty)) \ , \ \
H(\infty)+iK(\infty)=\exp(i\frac{\theta(\infty)}{2})
\ , \end{equation}
where $\theta(\infty)$ is an unknown function of $\xi$.
Therefore we choose
the boundary conditions
\begin{equation}
f_A'(\infty)=f_B'(\infty)=H'(\infty)=K'(\infty)=0
\ . \end{equation}
For the gauge function $\Phi$ we choose the boundary conditions
\begin{equation}
\Phi(0)=0 \ , \ \ \Phi'(\infty)=0
\ , \end{equation}
consistent with the boundary conditions
of the gauge and Higgs field functions.

\subsection{Results}

In the calculations presented
we have employed the gradient formalism
with constraint on the distance,
since it appeared numerically far more stable than
the explicit gradient approach.
Let us now discuss the sphaleron barrier
as obtained in the gradient approach.

{\it Barrier}

We first compare the gradient path
with the minimum energy path
for small Higgs boson masses.
In Fig.~1 we show
the energy as a function of the Chern-Simons number
for $M_H=M_W$
for the gradient path and the minimum energy path.
The minimum energy path barrier is steeper
with respect to the Chern-Simons number
than the gradient path barrier.
This picture reverts,
when we consider the energy as a function of the pathlength $l$,
defined by
\begin{equation}
l=\int_{vacuum}^{f} df -\int^{(bi)sphaleron}_{vacuum} df
\ . \end{equation}
Note, that the pathlength $l$ is shifted such that the sphaleron
or bisphaleron has $l=0$.
In Fig.~2 we show the energy
as a function of the pathparameter $l$
for $M_H=M_W$
along the gradient path and the minimum energy path.
With respect to the pathparameter $l$
the gradient barrier is steeper.

In Fig.~3 we present two configurations with the same
Chern-Simons number,
one along the gradient path
and one along the minimum energy path.
For the latter configuration
the asymptotic values of the functions
are already closer to the vacuum values.

In Fig.~4 we show the energy, obtained with the gradient method,
as a function of the Chern-Simons number
for $M_H=15M_W$.
Now there are three extrema of the energy functional,
the two degenerate bisphalerons and the symmetric sphaleron.
The right bisphaleron barrier is obtained from
the left one by the transformation
$N_{\rm CS} \rightarrow 1-N_{\rm CS}$
and $E \rightarrow E$.
All three barriers, the lower asymmetric bisphaleron barriers
and the higher symmetric sphaleron barrier,
are smooth in the gradient approach,
in contrast with the bifurcations encountered along
the extremal energy path.
Note, that the asymmetric bisphaleron barrier
culminates at the bisphaleron and has on one side
a very steep fall-off \cite{yaffe2}.
In Fig.~5 we show these energy barriers
as functions of the pathparameter $l$.

The effective mass $m$ is shown in Fig.~6
as a function of the Chern-Simons number
for $M_H=M_W$ and $M_H=15M_W$.
For small Higgs boson masses,
we find a smooth effective mass,
qualitatively similar to \cite{bitar1}.
But for large Higgs boson masses the effective mass
develops a sharp peak, when the Higgs field crosses zero
at spatial origin.
This point coincides with the symmetric sphaleron,
but not with the bisphaleron.
In the vicinity of this peak the potential energy
falls off steeply.
This steep fall-off occurs only on one side
(the side of the peak) of the bisphaleron,
but on both sides of the symmetric sphaleron.

By computing the distance of the (symmetric) sphaleron to
the vacuum we obtain an estimate of how good the paths are.
For $M_H=M_W$ the distance is 40, while the pathlength
of the gradient path is 46, and the pathlength of the minimum
energy path is 50.
Thus the minimum energy path has a longer pathlength
than the gradient path.

{\it Tunneling amplitude}

As a related criterion for the quality of a path
let us now consider
the associated semiclassical tunneling amplitude,
determined by $\exp(-R_0)$
\cite{banks1,banks2,bitar1,bitar2,noble,hsu}
\begin{equation}
R_0=\int_a^b d\lambda \sqrt{2m(\lambda)(V(\lambda)-E)}
\ , \end{equation}
where $V(\lambda)$ is the potential energy
and $E$ is the energy of the
classical turning points $a$ and $b$ \cite{bitar1,bitar2}.
(Note that the integral is independent of a reparametrization.)
Considering vacuum to vacuum transitions,
(with respect to the above metric, Eq.~(\ref{dis2}),)
the exponent $R_0$ of the tunneling amplitude
is a line integral along the path in configuration space,
\begin{equation}
R_0=\int_a^b \hat{d}f\  \sqrt{\frac{E_b(f)}{M_W}}
\ , \label{r0} \end{equation}
where
\begin{equation}
\hat d f = \lim_{\tilde f \rightarrow f} \hat d(f,\tilde f)
\ , \end{equation}
$V(\lambda)=E_{\rm b}$,
and $E=0$.
Employing the radial gauge in the calculations,
the tunneling amplitude is determined by
$R_0$ as given in Eq.~(\ref{r0}) with
$\hat d(f,\tilde f)$, Eq.~(\ref{dis2}),
replaced by
$d(f,\tilde f)$, Eq.~(\ref{dis3}),
where we now interpret the sequence of
configurations along the path
as gauge transforms of those configurations
for which $d(f,\tilde f)$ equals $\hat d(f,\tilde f)$.

For $M_H=M_W$ we find
$R_0=1.575$ and 1.836, in units of
$\frac{8\pi^2}{g^2}$,
for the gradient path and
the minimum energy path, respectively.
For $M_H=15 M_W$ we find
$R_0 =1.525$ (1.533), in units of
$\frac{8\pi^2}{g^2}$, for the bisphaleron barrier
(symmetric sphaleron barrier)
in the gradient approach.
In Fig.~7 we show $R_0$ as a function of the Higgs boson mass
in the gradient approach. We observe, that
$R_0 \approx 1.5$, in units of $8\pi^2/g^2$,
fairly independent of the Higgs boson mass,
and the bisphaleron transition rate is slightly higher
than the sphaleron transition rate.

\section{Level Crossing}

Let us now consider the fermionic level crossing phenomenon
along the sphaleron barrier.
We study the fermion mode in the background field
of the barrier as well as selfconsistently,
and compare the gradient path to the extremal energy path.

\subsection{Energy Functional}

To retain spherical symmetry
we consider only fermion doublets degenerate in mass.
The corresponding spherically symmetric ansatz for
the fermion eigenstates is the hedgehog ansatz,
\begin{equation}
q_L(\vec r\,,t) = e^{-i\omega t} M_W^{\frac{3}{2}}
\bigl[ G_L(r)
+ i \vec \sigma \cdot \hat r F_L(r) \bigr] \chi_{\rm h}
\ , \end{equation}
\begin{equation}
q_R(\vec r\,,t) = e^{-i\omega t} M_W^{\frac{3}{2}}
\bigl[ G_R(r)
- i \vec \sigma \cdot \hat r F_R(r) \bigr] \chi_{\rm h}
\ , \end{equation}
where the normalized hedgehog spinor $\chi_{\rm h}$
satisfies the spin-isospin relation
\begin{equation}
\vec \sigma \chi_{\rm h} + \vec \tau \chi_{\rm h} = 0
\ . \end{equation}
Under the residual gauge transformation, Eq.~\ref{gauge2},
the fermion functions transform as
\begin{eqnarray}
F_L+iG_L &\longrightarrow&  \exp(i\frac{\Theta}{2})(F_L+iG_L) \ , \nonumber\\
F_R+iG_R &\longrightarrow&  F_R+iG_R \ .
\end{eqnarray}

The fermionic energy functional reads
\begin{eqnarray}
 E_{\rm f} & = & 4\pi M_W \int^{\infty}_0 dx x^2
         \Bigl[  F_R'G_R-G_R'F_R+\frac{2}{x}F_RG_R
         +    F_L'G_L-G_L'F_L+\frac{2}{x}F_LG_L
\nonumber \\
&-& 2 \frac{1-f_A}{x}G_LF_L
+\frac{f_B}{x} (G_L^2-F_L^2)
+\frac{f_C}{2x} (G_L^2+F_L^2)
\nonumber \\
&+&      2\tilde{M}_F H(G_RG_L-F_RF_L)
       - 2\tilde{M}_F K(F_RG_L+F_LG_R)
\Bigr]
\ , \end{eqnarray}
where the fermion mass $M_F$ is expressed in units of $M_W$
\begin{equation}
\tilde{M}_F= M_F/M_W \nonumber
\ . \end{equation}

The fermion functions
need to be normalized.
When $N$ fermions occupy the eigenstate
the normalization condition is
\begin{equation}
         4\pi  \int^{\infty}_0 dx x^2
(G_R^2+F_R^2+G_L^2+F_L^2)=N
\ . \label{norm} \end{equation}

\subsection{Background Field Calculation}

Let us first consider the fermions in the background field
of the sphaleron barrier.
We find the set of coupled equations \cite{kb2,kb3,nk1}
\begin{equation}
\tilde \omega G_L - F'_L - \frac{2}{x}F_L
+\frac{1-f_A}{x} F_L
-\frac{f_B}{x} G_L
+\tilde M_F(-H G_R + K F_R) = 0
\ , \end{equation}
\begin{equation}
\tilde \omega F_L + G'_L
+\frac{1-f_A}{x} G_L
+\frac{f_B}{x} F_L
+\tilde M_F(H F_R + K G_R) = 0
\ , \end{equation}
\begin{equation}
-\tilde \omega G_R + F'_R + \frac{2}{x}F_R
+\tilde M_F(H G_L - K F_L) = 0
\ , \end{equation}
\begin{equation}
\tilde \omega F_R + G'_R
+\tilde M_F(H F_L + K G_L) = 0
\ , \end{equation}
where $\tilde \omega$ is the fermion eigenvalue $\omega$
in units of $M_W$
\begin{equation}
\tilde \omega = \frac{\omega}{M_W}
\nonumber \\
\ . \end{equation}

At the origin the fermion functions satisfy
the boundary conditions
\begin{equation}
F_R(0)=F_L(0)=0
\ , \end{equation}
and
\begin{equation}
G_R(0)=c_R \ , \ \ G_L(0)=c_L
\ , \end{equation}
where $c_R$ and $c_L$ are unknown constants,
subject to the normalization condition (\ref{norm}).
At infinity all fermion functions vanish
\begin{equation}
F_R(\infty)=F_L(\infty)=G_R(\infty)=G_L(\infty)=0
\ . \end{equation}

Let us now consider the case $M_H=M_W$.
The fermion eigenvalue
along the gradient path
is shown in Fig.~8
as a function of the Chern-Simons number
for the fermion masses $M_F=10 M_W$,
$M_F=M_W$ and $M_F=M_W/10$.
For comparison the fermion eigenvalue
along the minimum energy path is also shown \cite{kb2}.
For small fermion masses
the fermions are bound only
in the vicinity of the sphaleron.
Here we find qualitatively the same behaviour
of the fermion eigenvalue.
For heavier fermions the eigenmode reaches
the continua later along the gradient path
than along the minimum energy path.

For large values of the Higgs boson mass,
when the barrier culminates at the bi\-sphaleron,
the fermion eigenvalue in the gradient approach
is a monotonic function of the Chern-Simons number
as shown in Fig.~9 for $M_H=15M_W$
and $M_F=10 M_W$, $M_F=M_W$ and $M_F=M_W/10$.
This is in contrast with the extremal energy path,
where the bifurcations along the path
also lead to bifurcations of the fermion eigenvalue
\cite{kb3}.
Note, that the eigenvalue along the right bisphaleron barrier
is obtained by the transformation
$N_{\rm CS} \rightarrow 1-N_{\rm CS}$
and $\omega \rightarrow -\omega$.

In Fig.~10 we show the dependence of the zero eigenvalue
of the fermions on the Chern-Simons number
and on the fermion mass for the gradient path
and the extremal energy path for $M_H=15 M_W$.
Depending on the Higgs boson mass,
the zero mode approaches a limiting value
for large fermion masses in the gradient approach.
In contrast, along the extremal energy path
the zero mode occurs
for large fermion masses only at the sphaleron,
i.~e.~at $N_{\rm CS}=1/2$.
For small fermion masses the level crossing occurs
for both methods in the vicinity of
$N_{\rm CS}=1/2$ \cite{kb3,yaffe2}.

\subsection{Selfconsistent Calculation}

Let us now study the gradient path
over the sphaleron barrier in the presence of fermions.
We proceed analogously to our previous calculation
\cite{nk1}, but compute the barrier with the gradient method.
We arrive at the same set of equations for the fermion fields,
while we have to add the source terms
\begin{equation}
+g^2xF_LG_L
\ , \end{equation}
\begin{equation}
+\frac{1}{2}g^2x(G_L^2-F_L^2)
\ , \end{equation}
\begin{equation}
+\frac{g^2 \tilde M_F}{2}(G_RG_L-F_RF_L)
\ , \end{equation}
\begin{equation}
- \frac{g^2 \tilde M_F}{2}(F_RG_L+F_LG_R)
\ , \end{equation}
to the right hand side of the boson field equations for
$f_A$, $f_B$, $H$, and $K$, respectively.

Let us first consider small Higgs boson masses,
where only the sphaleron barrier exists.
As before,
when studying the fermion eigenmode
along the minimum energy path \cite{nk1},
we observe that
fermions with small masses have little influence
on the shape of the barrier,
while heavy fermions deform the barrier considerably.
However, for very large fermion masses,
the bifurcations, which we observed previously
along the extremal energy path,
are no longer present along the gradient path.
In the gradient approach
the barrier decreases monotonically
to both sides of the sphaleron,
as shown in Fig.~11 for $M_H=M_W$ and $M_F=75M_W$.

As before \cite{nk1}
we observe, that the fermion eigenvalue
deviates litte from the eigenvalue of
the background field calculation
for small fermion masses,
also for heavier fermions
the path does not lead to a free fermion solution
but to a bound state, a nontopological soliton.
The selfconsistent fermion eigenvalue
along the gradient barrier
is shown in Fig.~12
for a heavy fermion with $M_F=10 M_W$
for the Higgs boson mass $M_H=M_W$,
and compared to the eigenvalue of the
background field calculation.
In the selfconsistent calculation
the soliton is approached for $N_{\rm CS}\rightarrow 0$.

Let us now turn to large values of the Higgs boson mass,
where we expect two bi\-sphaleron barriers beside
the sphaleron barrier.
The presence of the fermions lifts the
degeneracy of the two bisphalerons
for finite fermion masses.
Considering now the total energy, consisting of the bosonic energy
and the fermion eigenvalue
as encountered along the path over the barrier,
we expect \cite{prob}, that the energy of the left bisphaleron
first increases as $E=E_{\rm b}+M_F$,
while the energy of the right bisphaleron
decreases as $E=E_{\rm b}-M_F$.
(The left sphaleron is encountered
along the barrier before the level crossing,
i.~e.~the fermion is still in the positive continuum,
while the right sphaleron is encountered after the level
crossing, i.~e.~the fermion is in the negative continuum
\cite{kb3}.)
At a critical value of the fermion mass
the fermion becomes bound,
$M_F^{cr} \approx 4 \sqrt{M_H/M_W-12}$ GeV
\cite{bk2}.
Then the energy of the left bisphaleron decreases,
while the energy of the right bisphaleron increases.
Interestingly, a bifurcation occurs
at a moderate value of the fermion mass.
The right bisphaleron merges with the sphaleron
at a critical value, beyond which
only the left bisphaleron solution exists.
This curious feature of the selfconsistent treatment
is demonstrated in Fig.~13 for $M_H=15 M_W$
(for one bound fermion).

The selfconsistent fermion eigenvalue
for a heavy fermion with $M_F=10 M_W$
for the Higgs boson mass $M_H=15 M_W$
along the left gradient bisphaleron barrier
is shown in Fig.~12.
Note, that this is the only selfconsistent barrier
for this fermion mass.
Here, the vacuum is approached for $N_{\rm CS}\rightarrow 0$,
because the soliton
exists only for higher fermion masses \cite{nk1}.

\section{Conclusion}

We have applied the gradient approach
to obtain the sphaleron barrier.
The gradient approach produces a path of steepest descent
with respect to a given metric.
A natural metric on the space of field configurations is
implied by the kinetic energy term of the Lagrangian \cite{met}.
We have formulated this metric
in a gauge invariant way \cite{fub}.

We have presented the formalism of the explicit
gradient approach and of the gradient approach with a
constraint on the distance.
For technical reasons
we have used the latter approach in the numerical
calculations.

Since the bifurcations along the extremal energy path
have largely motivated this study,
we have compared the sphaleron barrier obtained
along the gradient path
to the one along the extremal energy path.
For small values of the Higgs boson mass,
there is only one sphaleron solution,
the symmetric sphaleron \cite{km,kb1,yaffe}.
Here both approaches lead to a smooth barrier.
The gradient path barrier is steeper with
respect to the pathlength $l$, defined via
the metric,
while the minimum energy path is steeper
with respect to the Chern-Simons number.
But the semiclassically calculated tunneling amplitude,
$\sim\exp(-R_0)$, is bigger for the gradient path,
e.~g.~for $M_H=M_W$ $R_0$ is smaller by 10\% for the gradient path.

For Higgs boson masses larger than $\sim 1$ TeV
new asymmetric sphaleron solutions
with lower energy appear, the bisphalerons, \cite{kb1,yaffe}.
When bisphalerons exist, the extremal energy path
has bifurcations and culminates not at the
bisphaleron but at the symmetric sphaleron.
In contrast, the gradient approach leads
to smooth barriers, a lower asymmetric bisphaleron barrier
and a higher symmetric sphaleron barrier.
The asymmetric barrier has a steep fall-off on one side.
This fall-off is related to a peak in the effective mass,
when the Higgs field passes zero (at the origin).
The semiclassical tunneling amplitude
is fairly independent of the Higgs boson mass,
but slightly bigger along the bisphaleron barrier
than along the sphaleron barrier.

To exhibit the level crossing phenomenon
we have calculated the valence fermion mode
along the gradient approach barriers.
Since the barriers are smooth in the gradient approach,
also the fermion eigenvalue along the barriers
is smooth.
This is in contrast to the extremal energy path,
where the bifurcations of the barriers
were reflected in bifurcations of the fermion eigenvalue \cite{kb3}.
The fermion eigenvalue decreases monotonically
from the positive continuum to the negative continuum
along the gradient path
in the background field approximation,
even for large values of the Higgs boson mass.

When fermions are coupled selfconsistently to the boson fields
the fermion mass is of importance.
For small fermions masses there is hardly any change
with respect to the background field calculations,
while for heavy fermions the barriers deform
considerably.
Notably, for large Higgs boson masses,
two of the three barriers, the sphaleron barrier and the right
bisphaleron barrier, merge and disappear already for
moderate values of the fermion mass,
leaving as the only barrier the left bisphaleron barrier.

{\bf Acknowledegement}

We gratefully acknowledge discussions with Y. Brihaye,
B. Kleihaus, M. Wendel and L. Yaffe.

\newpage

\section{Figure Captions}

\indent{\bf Figure 1:}
The total energy (in TeV) is shown
as a function of the Chern-Simons number $N_{\rm CS}$
along the gradient path (solid)
and along the extremal path (dotted)
for $M_H=M_W$.

{\bf Figure 2:}
The total energy (in TeV) is shown
as a function of the pathparameter $l$
along the gradient path (solid)
and along the extremal path (dotted)
for $M_H=M_W$.

{\bf Figure 3:}
The gauge field functions $f_A$ and $f_B$ are shown
with respect to the dimensionless variable $x$
for the configurations with $N_{\rm CS}=1/2$
along the gradient path (solid)
and along the extremal path (dotted)
for $M_H=M_W$.

{\bf Figure 4:}
The total energy (in TeV) is shown
as a function of the Chern-Simons number $N_{\rm CS}$
along the symmetric sphaleron path (solid)
and along the asymmetric bisphaleron path (dotted)
in the gradient approach
for $M_H=15 M_W$.

{\bf Figure 5:}
The total energy (in TeV) is shown
as a function of the pathparameter $l$
along the symmetric sphaleron path (solid)
and along the asymmetric bisphaleron path (dotted)
in the gradient approach
for $M_H=15 M_W$.

{\bf Figure 6:}
The effective mass (in units of $1/M_W$) is shown
as a function of the Chern-Simons number $N_{\rm CS}$
in the gradient approach
along the symmetric sphaleron path
for $M_H=M_W$ (solid) and $M_H=15 M_W$ (dot-dashed),
and along the asymmetric bisphaleron path
for $M_H=15 M_W$ (dotted).

{\bf Figure 7:}
The transition amplitude $R_0$
(in units of $8\pi^2/g^2$) is shown
as a function of the Higgs boson mass $M_H$
(in units of $M_W$)
in the gradient approach
for the symmetric sphaleron path (solid)
and for the asymmetric bisphaleron path (dotted).

{\bf Figure 8:}
The fermion eigenvalue (in units of $M_F$) is shown
as a function of the Chern-Simons number $N_{\rm CS}$
in the background of the sphaleron barrier
for the fermion masses
$M_F=M_W/10$, $M_F=M_W$ and $M_F=10 M_W$
along the gradient path (solid)
and along the extremal path (dotted)
for $M_H=M_W$.

{\bf Figure 9:}
The fermion eigenvalue (in units of $M_F$) is shown
as a function of the Chern-Simons number $N_{\rm CS}$
in the background of the bisphaleron barrier
for the fermion masses
$M_F=M_W/10$, $M_F=M_W$ and $M_F=10 M_W$
along the gradient path (solid)
for $M_H=15 M_W$.

{\bf Figure 10:}
For the fermion zero mode
the dependence of the fermion mass (in units of $M_W$)
on the Chern-Simons number $N_{\rm CS}$ is shown
along the gradient path (solid)
and along the extremal path (dotted)
for $M_H=15 M_W$.

{\bf Figure 11:}
The energy (in TeV) is shown
as a function of the Chern-Simons number $N_{\rm CS}$
for the fermion mass $M_F=75 M_W$
in the selfconsistent calculation
along the gradient path (solid)
and along the extremal path (dotted)
for $M_H=M_W$.

{\bf Figure 12:}
The fermion eigenvalue (in units of $M_F$) is shown
as a function of the Chern-Simons number $N_{\rm CS}$
in the gradient approach
along the sphaleron barrier for $M_H=M_W$
and along the bisphaleron barrier for $M_H=15 M_W$
for the fermion mass $M_F=10 M_W$
in the background field calculation (dotted)
and in the selfconsistent calculation (solid).

{\bf Figure 13:}
The energy (in TeV),
including the fermion eigenvalue, is shown
as a function of the fermion mass $M_F$ (in units of $M_W$)
in the selfconsistent calculation
for the left bisphaleron (dotted),
for the right bisphaleron (solid),
and for the sphaleron (dot-dashed)
for $M_H=15M_W$.


\begin{thebibliography}{000}

\bibitem{hooft}
G. 't Hooft,
Symmetry breaking through Bell-Jackiw Anomalies,
Phys. Rev. Lett. 37 (1976) 8.

\bibitem{ring1}
A. Ringwald,
Rate of anomalous electroweak baryon and lepton number
violation at finite temperature,
Phys. Lett. B201 (1988) 510.

\bibitem{ssc}
M. Mattis, and E. Mottola, eds.,
``Baryon Number Violation at the SSC?'',
World Scientific, Singapore (1990).

\bibitem{man}
N.~S. Manton,
Topology in the Weinberg-Salam theory,
Phys. Rev. D28 (1983) 2019.

\bibitem{km}
F.~R. Klinkhamer, and N.~S. Manton,
A saddle-point solution in the Weinberg-Salam theory,
Phys. Rev. D30 (1984) 2212.

\bibitem{krs}
V.~A. Kuzmin, V.~A. Rubakov, and M.~E. Shaposhnikov,
On anomalous electroweak baryon-number non-conservation
in the early universe,
Phys. Lett. B155 (1985) 36.

\bibitem{mcl1}
P. Arnold, and L. McLerran,
Sphalerons, small fluctuations, and baryon-number violation
in electroweak theory,
Phys. Rev. D36 (1987) 581.

\bibitem{mcl2}
P. Arnold, and L. McLerran,
The sphaleron strikes back: A response
to objections to the sphaleron approximation,
Phys. Rev. D37 (1988) 1020.

\bibitem{mcl3}
L. Carson, X. Li, L. McLerran, and R.-T. Wang,
Exact computation of the small-fluctuation determinant
around a sphaleron,
Phys. Rev. D42 (1990) 2127.

\bibitem{kolb}
E.~W. Kolb, and M.~S. Turner,
``The Early Universe'',
Addison-Wesley Publishing Company, Redwood City (1990).

\bibitem{akiba}
T. Akiba, H. Kikuchi, and T. Yanagida,
Static minimum-energy path from a vacuum to a sphaleron
in the Weinberg-Salam model,
Phys. Rev. D38 (1988) 1937.

\bibitem{bgkk}
Y. Brihaye, S. Giler, P. Kosinski, and J. Kunz,
Configuration space around the sphaleron,
Phys. Rev. D42 (1989) 2846.

\bibitem{kb1}
J. Kunz, and Y. Brihaye,
New sphalerons in the Weinberg-Salam theory,
Phys. Lett. B216 (1989) 353.

\bibitem{yaffe}
L.~G. Yaffe,
Static solutions of SU(2)-Higgs theory,
Phys. Rev. D40 (1989) 3463.

\bibitem{nohl}
C.~R. Nohl,
Bound-state solutions of the Dirac equation in extended
hadron models,
Phys. Rev. D12 (1975) 1840.

\bibitem{boguta}
J. Boguta, and J. Kunz,
Hadroids and sphalerons,
Phys. Lett. B154 (1985) 407.

\bibitem{ring2}
A. Ringwald,
Sphaleron and level crossing,
Phys. Lett. B213 (1988) 61.

\bibitem{kb2}
J. Kunz, and Y. Brihaye,
Fermions in the background of the  sphaleron barrier,
Phys. Lett. B304 (1993) 141.

\bibitem{dia}
D. Diakonov, M. Polyakov, P. Sieber, J. Schaldach and K. Goeke,
Fermion sea along the sphaleron barrier,
Phys. Rev. D49 (1994) 6864.

\bibitem{kb3}
J. Kunz, and Y. Brihaye,
Level crossing along sphaleron barriers,
Phys. Rev. D50 (1994) 1051.

\bibitem{kkb1}
B. Kleihaus, J. Kunz, and Y. Brihaye,
The electroweak sphaleron at physical mixing angle,
Phys. Lett. B273 (1991) 100.

\bibitem{kkb2}
J. Kunz, B. Kleihaus, and Y. Brihaye,
Sphalerons at finite mixing angle,
Phys. Rev. D46 (1992) 3587.

\bibitem{nk1}
G. Nolte, and J. Kunz,
The sphaleron barrier in the presence of fermions,
Phys.Rev. D48 (1993)  5905.

\bibitem{dhn}
R.~F. Dashen, B. Hasslacher, and A. Neveu,
Nonperturbative methods and extended-hadron models
in field theory. III.~Four-dimensional non-abelian models,
Phys. Rev. D12 (1974) 4138.

\bibitem{klink}
F.~R. Klinkhamer,
Sphalerons, deformed sphalerons and configuration space,
Phys. Lett. B236 (1990) 187.

\bibitem{bk1}
Y. Brihaye and J. Kunz,
Normal modes around SU(2) sphalerons,
Phys. Lett. B249 (1990) 90.

\bibitem{met}
This choice of metric is within the gradient approach
consistent with the requirement of a maximal tunneling
rate.

\bibitem{banks1}
T. Banks, C.~M. Bender, and T.~T. Wu,
Coupled anharmonic oscillators. I. Equal-mass case,
Phys. Rev. D8 (1973) 3346.

\bibitem{banks2}
T. Banks, and C.~M. Bender,
Coupled anharmonic oscillators. II. Unequal-mass case,
Phys. Rev. D8 (1973) 3366.

\bibitem{bitar1}
K. Bitar, and S.-J. Chang,
Vacuum tunneling of gauge theory in Minkowsky space,
Phys. Rev. D17 (1978) 486.

\bibitem{bitar2}
K. Bitar, and S.-J. Chang,
Vacuum tunneling and fluctuations around a most probable escape path,
Phys. Rev. D18 (1978) 435.

\bibitem{noble}
R. J. Noble,
Quantum-field transition rates at finite temperature,
Phys. Rev. D20 (1979) 3179.

\bibitem{hsu}
S. Hsu,
On tunneling at finite energies and temperatures,
Phys. Lett. B294 (1992) 77.

\bibitem{fub}
This metric is a generalization of the
Fubini-Study metric \cite{pati} to local gauge theory,
which is defined for quantum
mechanical wavefunctions, which are physically equivalent  under
a global phase transformation.

\bibitem{pati}
A.~K. Pati,
Relation between ``phases'' and ``distance'' in quantum evolution,
Phys. Lett. A159 (1991) 105.

\bibitem{yaffe2}
Similar results were obtained by L.~G. Yaffe, private communication.

\bibitem{bk2}
Y. Brihaye and J. Kunz,
Sphaleron, Fermions and Level Crossing in the Electroweak Model,
ICHEP94, Glasgow, July 1994.

\bibitem{prob}
Due to the finite radius, we had numerical difficulties
to obtain accurate (free fermion) solutions
for small fermion masses.

\end{thebibliography}
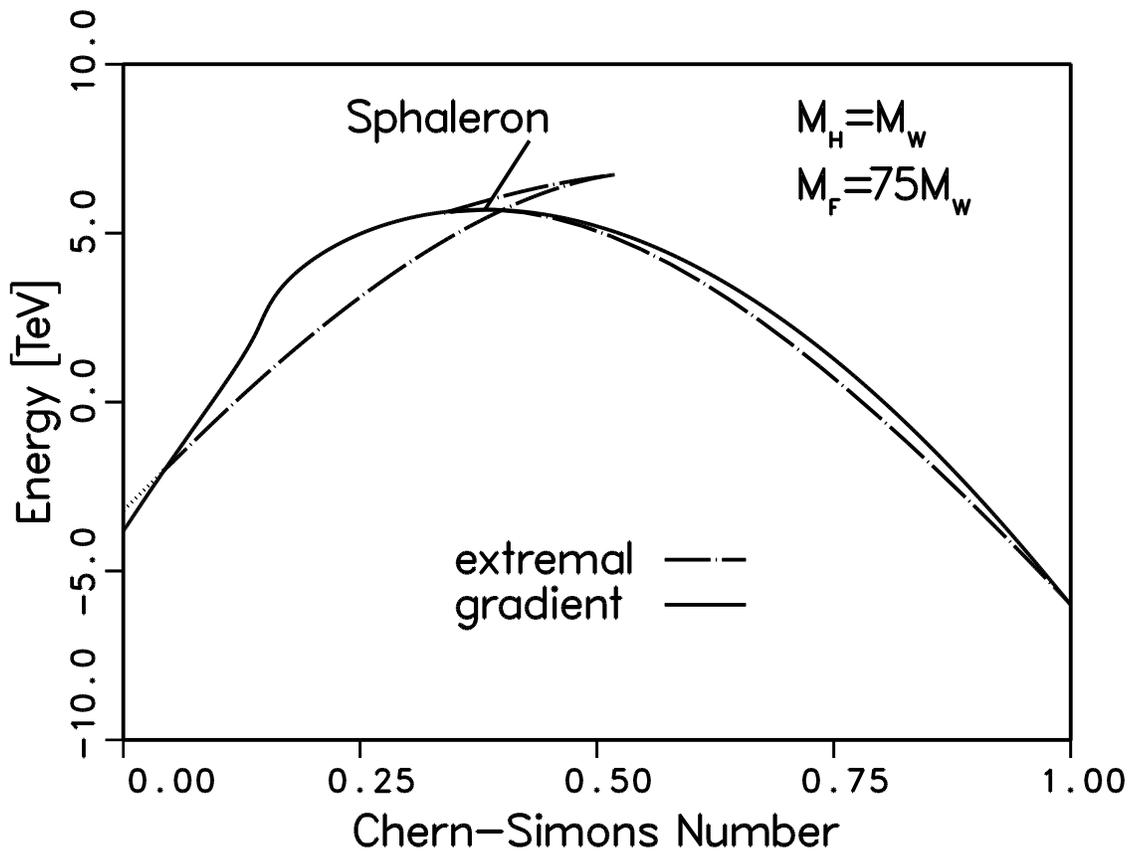
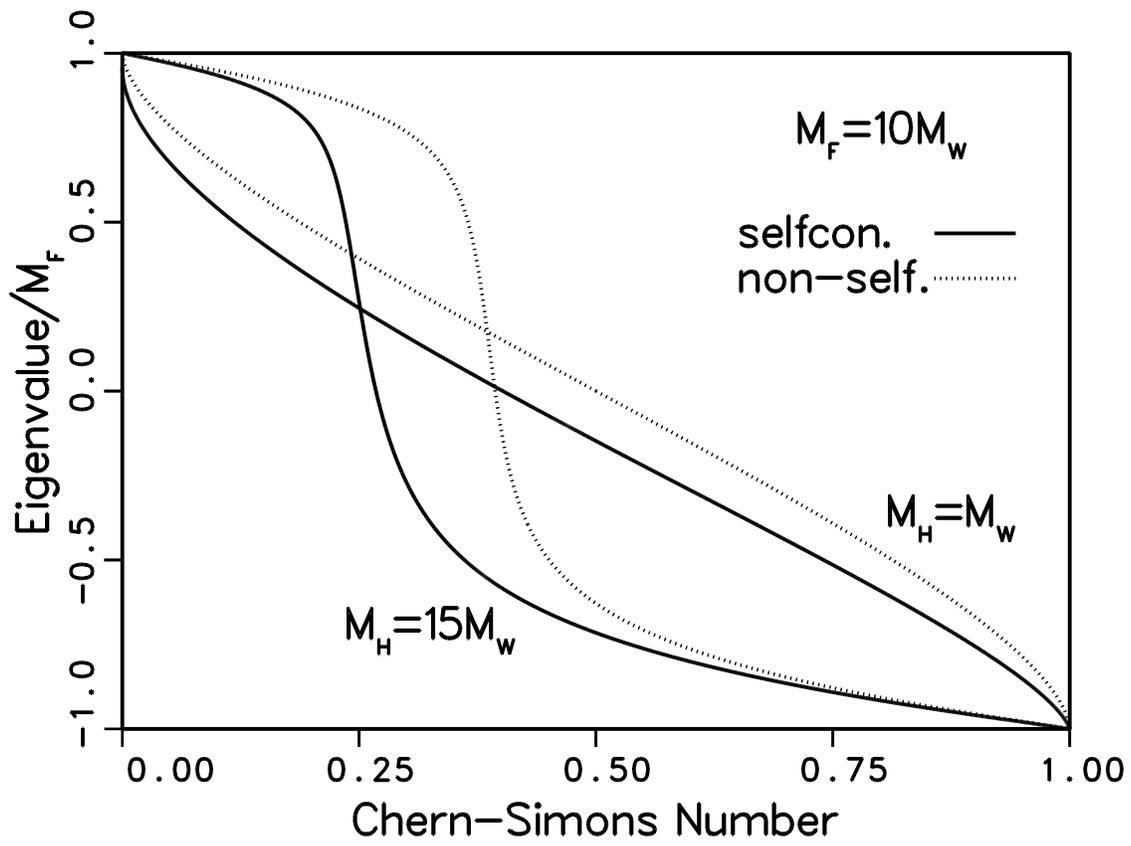
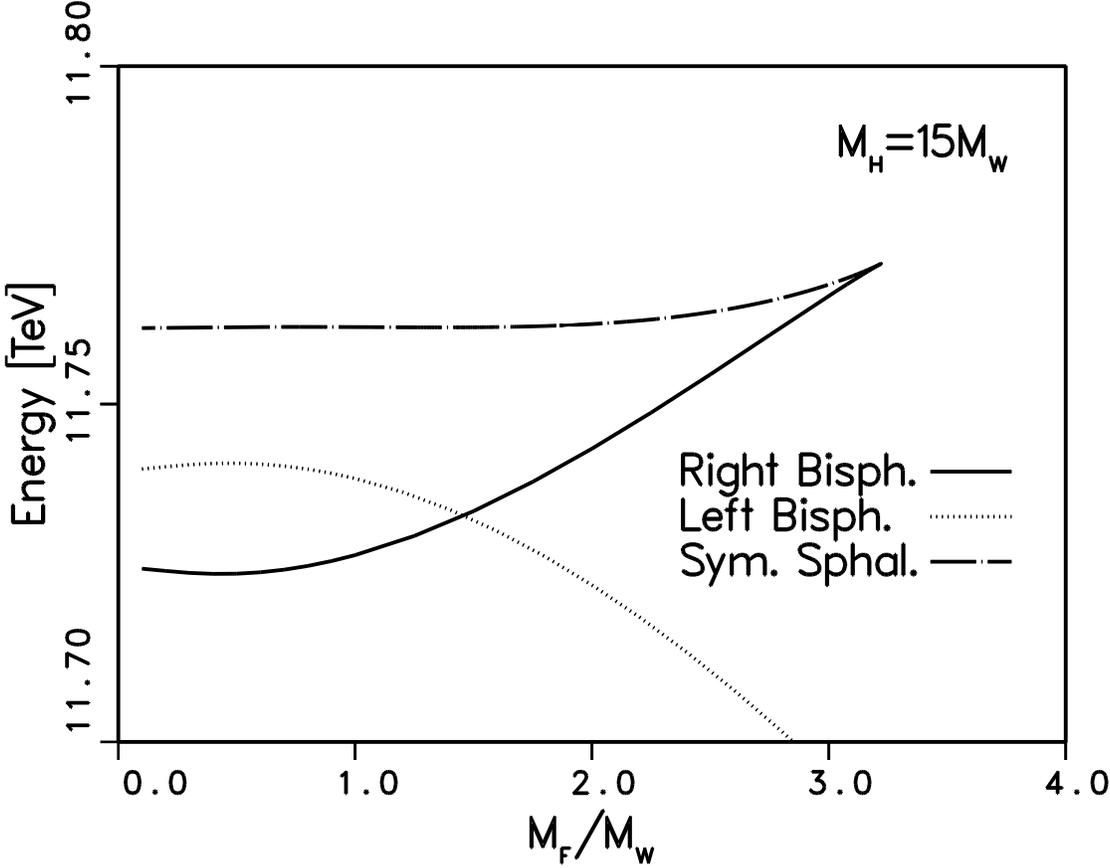
\end{document}